\documentclass[sigconf]{acmart}
\AtBeginDocument{%
  }

\copyrightyear{2026}
\acmYear{2026}
\setcopyright{cc}
\setcctype{by-nc-nd}
\acmConference[CHI '26]{Proceedings of the 2026 CHI Conference on Human Factors in Computing Systems}{April 13--17, 2026}{Barcelona, Spain}
\acmBooktitle{Proceedings of the 2026 CHI Conference on Human Factors in Computing Systems (CHI '26), April 13--17, 2026, Barcelona, Spain}
\acmPrice{}
\acmDOI{10.1145/3772318.3791524}
\acmISBN{979-8-4007-2278-3/2026/04}

\usepackage{multirow}
\begin{document}

\title{Designing with Medical Mistrust: Perspectives from Black Older Adults in Publicly Subsidized Housing }

\author{Cynthia M Baseman}
\affiliation{%
  \institution{Georgia Institute of Technology}
  \city{Atlanta}
  \state{Georgia}
  \country{USA}
}
\email{cbaseman3@gatech.edu}
\orcid{0009-0004-8507-438X}

\author{Reeda Shimaz Huda}
\affiliation{%
  \institution{Georgia Institute of Technology}
  \city{Atlanta}
  \state{Georgia}
  \country{USA}}
\email{rhuda8@gatech.edu}
\orcid{0009-0006-4495-0413}

\author{Rosa I Arriaga}
\affiliation{%
  \institution{Georgia Institute of Technology}
  \city{Atlanta}
  \state{Georgia}
  \country{USA}}
\email{arriaga@cc.gatech.edu}
\orcid{0000-0002-8642-7245}

\renewcommand{\shortauthors}{Baseman et al.}

\begin{abstract}
Despite increasing interest in culturally-sensitive health technologies, medical mistrust remains largely unexplored within human-centered computing. Considered a social determinant of health, medical mistrust is the belief that healthcare providers or institutions are acting against one's best interest. This is a rational, protective response based on historical context, structural inequities, and discrimination. To center race-based medical mistrust and the lived experiences of Black older adults with low income, we conducted interviews within publicly subsidized housing in the Southern United States. Our reflexive themes describe community perspectives on health care and medical mistrust, including accreditation and embodiment, skepticism of financial motivations, and the intentions behind health AI. We provide a reflective exercise for researchers to consider their positionality in relation to community engagements, and reframe our findings through Black Feminist Thought to propose design principles for health self-management technologies for communities with historically grounded medical mistrust. 
\end{abstract}

\ccsdesc[500]{Human-centered computing~Empirical studies in HCI}
\ccsdesc[500]{Social and professional topics~Race and ethnicity}
\ccsdesc[300]{Social and professional topics~Seniors}

\keywords{Medical Mistrust, Health, Race, Black or African American, Older Adults, Human-Centered Design, Health Equity}

\maketitle

\section{Introduction}

Whether flourishing or declining, physical or mental, health does not occur in a vacuum. The social and structural drivers of health are critical environmental, social, cultural, and infrastructural factors that influence an individual's health \cite{solar2010conceptual,brown2023future}. This more social-ecological understanding of health has motivated research promoting positive health practices and outcomes given the lived experiences of individuals and their larger environment. In fact, health professional education increasingly includes the social drivers of health and cultural competency education within curricula \cite{brottman2020toward,nonyel2021conceptualizing}. 

The human-computer interaction (HCI) community, too, has shown growing interest in designing health self-management technologies that appreciate the larger environmental and sociocultural factors  impacting users' daily lives. In particular, there is interest in how these factors influence perceptions of, and trust in, health technologies. Much HCI research distills design implications to increase the trust users place in specific health technologies (e.g., health chatbots \cite{li2025customizable,wang2025understanding,harrington2023trust,sun2025interface}) given their identity, experiences, and sociocultural background. Other researchers engage in community-centric approaches, working with specific racial or cultural communities to design culturally sensitive health chatbots \cite{deva2025kya,harrington2022s,o2020community} and mobile/digital health \cite{baseman2025we,claisse2024understanding,oguamanam2023intersectional,o2025empowering,vigil2021integrating}. Studies that center the perspectives of historically marginalized communities in health therefore exist, but they are few in number. In particular, a social determinant of health \cite{shukla2025medical} that has not yet been explored in depth by the HCI community is medical mistrust experienced by Black or African American communities in the United States.

Medical mistrust is not simply a lack of trust in the healthcare system. It can be defined as the belief that healthcare providers, systems, or institutions are acting against one's best interest or well-being \cite{hall2001trust,armstrong2008differences}. Medical mistrust is complex and can arise from a combination of historical context, structural inequities, and daily experiences of discrimination \cite{jaiswal2019towards,souvatzi2024trust}.  In fields such as medicine and public health, medical mistrust is widely discussed due to its implications for patient health outcomes  \cite{souvatzi2024trust,jaiswal2019towards}. For example, institutional and medical mistrust led to higher COVID-19 vaccine hesitancy within Black communities \cite{laurencin2021addressing,sekimitsu2022exploring}. Much literature on medical mistrust, however, positions medical mistrust as a ``cultural barrier'' or attributes it to “conspiracy theories,” racist assumptions that undermine the validity of this mistrust \cite{jaiswal2019towards}. Ethical and equitable health technology design must dispel this deficit narrative, and instead accept that medical mistrust \cite{benkert2019ubiquitous} and associated skepticism of novel health technologies \cite{harrington2023trust,lee2021included} are rational and protective responses due to a history of injustice and inequity. 

This paper offers, to our knowledge, the first concerted comm\-unity-engaged study to specifically center race-based medical mistrust within human-centered computing. This work asks how researchers can design health self-management technologies not to ignore medical mistrust nor to determinedly attempt to decrease it, but rather to validate this reasonable, protective response to build bridges to a more equitable future. To explore this question, we leverage our team's community partnerships with two publicly subsidized apartment buildings within a city in the Southern United States. As these community sites predominantly serve Black older adults with low income, our work centers the perspectives and lived experiences of an intersectionally marginalized community. We had two key research goals. First, anchored in lived experiences, we aimed \textbf{to better understand the perspectives} on doctors, the healthcare system, and medical mistrust held by community members living with diabetes. Second, we aimed to further \textbf{contextualize mistrust through a scenario of a health AI technology} for diabetes self-management.

This work provides three key contributions \textbf(C1-C3). \textbf{First,} in Section \ref{findings}, we provide eight themes distilled using reflexive thematic analysis \cite{braun2019reflecting,braun2023toward}, providing insights into medical mistrust and community health experiences (C1). These eight reflexive themes include skepticism regarding doctors' financial motivations, negotiations between trusting accreditation and embodied experiences, the value of taking personal accountability in one's health, and intentions behind health AI systems. \textbf{Second,} in Section \ref{reflection} we reflect on our study and how medical mistrust should impact an HCI researcher's approach to design, offering a reflective exercise for researchers to consider their positionality in relation to community engagements (C2). We discuss the tension between bolstering community agency and reducing the burden placed upon historically marginalized communities. \textbf{Third}, we reframe our findings through Black Feminist Thought \cite{collins1990black, collins2000black}. Through the lens of structural, disciplinary, hegemonic, and interpersonal power, there emerge seven principles for researchers to consider when designing health self-management technologies (C3). The proposed design framework offers insights into how to design more ethical and equitable health self-management technologies taking into consideration the historical context of medical mistrust, and supporting advocacy and resistance through design.

Medical mistrust is frequently associated with Black or African American communities due to the unique system of exploitation and experimentation within the U.S. medical system (see Section \ref{history_positionality}). Medical mistrust is also known to be prevalent, however, within any community that has faced inequities or discrimination by medical institutions, including but not limited to: other communities of color, women, LGBTQIA+ communities, and individuals with disabilities \cite{ho2022medical,wegner2025understanding,jaiswal2019whose}. While we center the lived experiences of two communities of Black older adults with low income, we provide higher level insights into health technology that may be transferable  \cite{soden2024evaluating} to other communities that experience medical mistrust, and the role of HCI researchers in moving towards health equity. 

\textbf{Content Warning:} This paper includes discussion of medical experimentation and other injustices perpetrated by doctors and medical institutions, as well as depictions of negative health experiences including deaths of loved ones.

\section{History \& Positionality: On Medical Mistrust}
\label{history_positionality}
We first provide a brief overview of the historical context of medical mistrust in Black, African American, and/or African communities in the United States. We then provide a statement to disclose our positionality and perspectives as they relate to our participants and our ongoing community engagement.

\subsection{The Complexities of Medical Mistrust in the United States}
\label{history}

Shukla et al. define medical mistrust as ``a social determinant of health fueled by an individual’s fear of harm and exploitation, presenting as suspicion of and lack of confidence in healthcare providers and systems, [that] is experienced at both the interpersonal, intergenerational, and institutional levels, reinforced by historical precedents and ongoing structural racism and systemic inequalities'' \cite{shukla2025medical}. Medical mistrust has been studied in fields including medicine, sociology, and public health due to its broad implications for patient outcomes \cite{souvatzi2024trust,jaiswal2019towards}.  Suspicion of healthcare professionals, racial inequities, and a lack of support may all influence the mistrust of Black patients \cite{cueva2024medical}. Much recent work has focused specifically on the impacts of medical mistrust and institutional distrust on COVID-19 vaccine skepticism  or distrust within Black communities \cite{laurencin2021addressing,sekimitsu2022exploring,smith2022investigation,ash2021predictors,madorsky2021vaccine}. Previous literature, however, reveals patient strategies of mitigating mistrust including empowerment through education and trusting one's intuition \cite{gagnon2025characteristics}. Below we provide some brief sociohistorical context to better situate medical mistrust and our work. 

Many readers may be familiar with the infamous Tuskegee Syphilis Study \cite{brandon2005legacy}. From 1932 to 1972, the U.S. Public Health Service engaged Black men in syphilis experiments, monitored them for many years, and withheld treatment (even after the widespread availability of penicillin) to determine the effects of untreated syphilis \cite{jaiswal2019towards}. The Tuskegee Syphilis Study is widely cited as a central catalyst of medical mistrust within Black communities \cite{brandon2005legacy,jaiswal2019towards}. Although this was undeniably an unethical study, it is not an anomaly but rather is  indicative of a pattern: There is an extensive history of experimentation, abuse, and injustice perpetrated by United States medical institutions and professionals against Black communities, especially in the Southern United States\footnote{Interested readers might consider reading \textit{Medical apartheid: The dark history of medical experimentation on Black Americans from colonial times to the present} by Harriet A. Washington \cite{washington2006medical}. This history was incredibly helpful for our authors in approaching this work.}. A few examples include:  medical experiments performed on enslaved persons (e.g., those performed by ``father of modern gynecology'' Marion Sims), the disproportionate use of Black cadavers in medical schools (facilitated not only by the illegal procurement of cadavers, but also trade from the South to Northern medical schools), and eugenics efforts (e.g., ``Mississippi appendectomy'' forced sterilization)  \cite{washington2006medical}. 

Sixty years ago (when almost all of this study's participants were children), hospitals were still legally segregated in the United States. Relatively wide desegregation began not directly due to the Civil Rights Act of 1964, but rather due to the  introduction of Medicare and Medicaid in 1965: Hospitals resisting desegregation would be charged higher costs \cite{marquez2023call}. United States hospitals were finally officially desegregated, but the lasting impacts of this de jure segregation, as well as persistent de facto racial segregation, are partially responsible for inequities of health insurance, resource distribution, and ultimately health disparities \cite{largent2018public,marquez2023call,carrasquillo2025informal}. To this day, there exist racial disparities in diabetes-related amputations: Black patients are more likely to receive an amputation rather than a procedure to salvage the limb \cite{lobo2023regional,stapleton2018variation}.

Medical mistrust can also arise (or grow) due to the ``daily experiences of racism, stigma, and discrimination'' faced by historically marginalized communities \cite{jaiswal2019towards}. Although some institutional barriers to health care faced by Black communities have been mitigated (e.g., de jure segregation), these communities still face ongoing discrimination within clinical settings, ranging from microaggressions, to being treated without respect or dignity by health professionals, to overt verbal insults \cite{brown2024discriminatory,hall2022experiences,garrett2024black}. In addition, health care is impacted by the legacy and current manifestations of racism within social drivers of health such as economic stability \cite{braveman2014social}, digital redlining  \cite{wang2024digital}, and educational attainment  \cite{braveman2014social}. Given the depth and breadth of this history, previous work has argued that there is a crucial need to foster better understandings of medical mistrust including the historical and ongoing systemic injustices within the U.S. healthcare system \cite{jaiswal2019towards}.

\subsection{Positionality}
\label{our_positionality}

We include our positionality statement here to better illustrate researcher identity as it relates to this sociohistorical context. Our three authors are human-centered computing researchers with backgrounds in engineering, computer science, and psychology. The first author identifies as white, queer, and transgender. The second author identifies as a South Asian Muslim woman. The third author identifies as a Latina researcher. Our team differs in the nuances and extent to which we are critical of medical institutions and experience medical mistrust. Women, LGBTQIA+ folks, ethic/racial minorities, and Muslim individuals in the U.S. (and especially those with intersecting marginalized identities) all may experience medical mistrust due to a history of exclusion and/or discrimination \cite{ho2022medical,wegner2025understanding,jaiswal2019whose}. We do not attempt to equate our experiences of medical mistrust to those experienced by Black Americans, as one cannot overstate the historical specificity of slavery and its extensive legacy. Still, we believe that having identity characteristics associated with medical mistrust made us more ready to listen to community members and believe in the validity of their lived experiences. 

\subsubsection{Project History}
\label{[project_history]}

This interview study took place within a broader multi-year partnership with the community sites (described in Section \ref{method_partnerships}). Our team's overall goal is to design more useful, usable, and equitable self-monitoring technologies for diabetic foot disease through community tailoring. Communities of color and those with low income are disproportionately impacted by diabetes and its complications. Diabetic foot ulcers, a life-threatening complication that frequently leads to infection and amputation, are no exception: Patients of color are more likely to undergo a lower-limb amputation \cite{lobo2023regional,stapleton2018variation, brennan2022association,lavery1996variation}. Previous work has shown the benefits of technological support for self-monitoring diabetic foot ulcers, especially utilizing computer vision  \cite{swerdlow2023initial,baseman2025intelligent}; however, gaps of digital literacy and technological familiarity could complicate these opportunities \cite{bondu2024accessible,stowell2018designing}, suggesting the need for community-tailored technologies. Our multi-year engagement aims to involve community members throughout each stage of design, driving a technology designed explicitly for their community. 

Two additional considerations arose throughout our engagements with the community sites. First, community members began to voice indications of mistrust. Medical mistrust sometimes increased the use of self-management technologies, but also interacted with distrust of the institutions creating these technologies. Second, there is evidence that state-of-the-art ulcer recognition models provide worse performance on people of color \cite{baseman2025towards}, suggesting that a critical or protective response to these technologies may be beneficial. These considerations drive the need to better understand and contextualize mistrust, in order to inform our approach to designing for/with this community.

\section{Related Work}

We now highlight three avenues of recent related work. First, we discuss the study of trust, skepticism, and distrust of health technologies (Section \ref{RW_trust_health}). We then take a step deeper into institutional distrust and medical mistrust in HCI, describing the distinction between distrust and \textit{mis}trust (Section \ref{RW_institutional_distrust_MM}). Finally, as a study of medical mistrust is dependent on access to situated, lived experiences, we provide an overview of the culturally sensitive and/or community-engaged design of health technologies (Section \ref{RW_community_based_health}).

\subsection{Trust \& Skepticism in Health AI Technologies}
\label{RW_trust_health}

Although this work specifically centers \textit{mis}trust, we first review the existing body of literature on trust. In recent years, HCI researchers have investigated trust, distrust, and skepticism of human-AI interactions \cite{ueno2022trust} and technologies ranging from service robots \cite{garcia2025being}, to civic technologies \cite{corbett2021designing}, to social media \cite{zhang2023we}. Other work has the goal of calibrating trust in AI systems based on the actual capabilities and limitations of the system \cite{metzger2024empowering,bo2025rely,schoeffer2024explanations}, driving ``appropriate reliance'' \cite{lee2004trust}. HCI for health is no exception to this focus on trust, and in fact, the importance of centering trust as an ethical principle may be even greater in health contexts \cite{swinger2025there}.

Recent explorations of trust in novel health technologies have provided design implications for system design. Literature on health-related conversational agents, for example, shows that user trust can be impacted by factors including: perceived moral agency \cite{wester2024chatbot}, ability to personalize the chatbot \cite{li2025customizable}, severity of symptoms \cite{wang2025understanding}, and the racial, gender, and age features of an embodied agent  \cite{harrington2023trust}. Further, trust in health information applications may be influenced by interface usability \cite{sun2024trust,sun2025interface}. Similar work on voice assistants has revealed that compared to general-purpose voice assistants, users' trust in health-related voice assistants places a larger emphasis on perceived usefulness and privacy considerations \cite{zhan2024healthcare}. 

A complement to this work on trust, other literature has explored attitudes of skepticism towards novel health technologies. This work especially centers older adults due to the ``digital divide,'' the widening gaps in access and effective use of technologies \cite{marimuthu2022challenging}. Older adults may be seen as ``critical adopters'' \cite{barros2021circumspect} of technology: They may approach a technology with skepticism if not knowledgeable about how it works  \cite{barros2021circumspect} and/or if the technology is misaligned with their values \cite{knowles2018older}. Recent work has revealed older adults' skepticism of personal health data when unfavorable (e.g., depicting low levels of physical activity) \cite{caldeira2023compare}, and doubt regarding information sources and intentions behind health technologies \cite{harrington2023trust}. 

While there is a large body of work exploring user trust in health technologies, much of this work has been artifact-specific, with a goal of providing concrete design recommendations to increase a technology's perceived trustworthiness. Less frequently does the literature explicitly engage with higher-level institutional and systemic factors that influence users' trust in health technologies. As Stowell et al. note, focusing solely on health outcomes (or, we could extend the claim, to technology trustworthiness) can neglect community, environmental, or institutional factors of health inequities \cite{stowell2018designing}. As we discuss further in the next section, we offer insights into health technology design taking into consideration medical mistrust as a  broader context of lived experience.

\subsection{Institutional Distrust \& Medical Mistrust in HCI}
\label{RW_institutional_distrust_MM}

Distrust of institutional actors can heavily influence the perspectives held by individuals excluded or marginalized by those institutions. Institutional distrust may shape their perceptions of the role future technologies should play in society, for example heightening considerations of privacy and power \cite{martinez2024engaging}. Some HCI research outside of health has therefore centered institutional distrust, especially of financial institutions, academic institutions, and online platforms. HCI work in financial technologies has viewed Cash App as an avenue Black communities turn to because of distrust of financial institutions \cite{cunningham2022cost}, and other work identified culturally situated factors influencing distrust of NFT projects \cite{cao2025jade}. Others have reflected on how institutional distrust of research institutions impacts relationships with community sites, and especially how to build trust and foster more equitable partnerships \cite{harrington2019deconstructing,baseman2025we,le2015strangers}. There is also recent literature centering online information. Institutional distrust in online platforms can both influence youth preferences for resolution after online harassment \cite{schoenebeck2021youth} and raise challenges for recruiting older adults given wariness of scams \cite{ankenbauer2025right}. In addition, due to threats of misinformation and anti-institution sentiments during the COVID-19 pandemic, much recent work has focused specifically on online health information \cite{zhang2022shifting,efstratiou2024here}, including for Black and Latinx communities (for which mis- or distrust of information may be due to historical inequities) \cite{chen2024we}. 

The present work specifically explores medical \textit{mis}trust. This is distinct from institutional distrust, although medical mistrust frequently co-occurs or is deepened/sustained by institutional distrust. More than an ``absence of trust'' \cite{jaiswal2019towards}, medical mistrust is a belief that healthcare providers, systems, or institutions are acting \textit{against one's best interest or well-being} \cite{hall2001trust,armstrong2008differences}. It can therefore arise from a combination of historical context, structural inequities, and daily experiences of discrimination (including the interpersonal) \cite{jaiswal2019towards}.  There is a small but growing body of HCI research investigating the impacts of medical mistrust or a distrust of medical institutions on Black communities' perceptions of health technologies. In particular, medical mistrust may translate into a skepticism of health technologies, including the researchers' intention behind the technology \cite{harrington2023trust,lee2021included}. Perhaps the most widely cited HCI paper on ``cultural mistrust,'' Lee and Rich show that U.S. Black and African Americans who experience higher medical mistrust may perceive health AI to be as untrustworthy as human medical professionals \cite{lee2021included}. Reflecting on ethical methodology, Baseman et al. urged HCI researchers to avoid health equity tourism, as this could contribute to medical mistrust within under-resourced communities \cite{baseman2025we}. Kim et al. found that some participants hoped a health chatbot could be a ``trustworthy voice of the community,'' providing information about COVID-19 vaccine safety \cite{kim2022designing}. Similarly, Chen et al. investigated how digital platforms can better support trusted community messengers in addressing COVID-19 vaccine hesitancy in Black and Latinx communities \cite{chen2024we}. 

Research on institutional distrust, ranging from financial to academic, makes evident that sociohistorical context influences perceptions of technologies. Very few HCI studies to date, however, have explored institutional distrust of healthcare institutions, and even fewer have explicitly discussed medical \textit{mis}trust. Through semi-structured interviews within community sites of Black older adults with low income, we provide insights into their perspectives of health care and medical mistrust (Section \ref{findings}). The nuance of medical mistrust (i.e., historical grounding and a lack of beneficence) may impact the implications for design and ethical methodologies in HCI. As this mistrust is based in situated lived experiences, a study of medical mistrust requires direct engagement with these communities to better understand their perspectives. We therefore devote the next section to community-engaged design.

\subsection{Community-engaged Design of Health Technologies}
\label{RW_community_based_health}

Continuing calls in HCI emphasize the importance of historical context and intersectional identity \cite{erete2018intersectional}, especially within health  \cite{harrington2022examining,harrington2018informing,stowell2018designing}. In recent years, HCI researchers have leveraged community-centric approaches to design more culturally tailored health technologies, such as: chatbots for reproductive health information for underserved women in India \cite{deva2025kya}, digital mental health applications for Black women  \cite{o2025empowering,oguamanam2023intersectional}, behavioral mobile health for Native American youth \cite{vigil2021integrating}, and mobile health to support pregnant Roma women  \cite{claisse2024understanding}. Health contexts offer an opportunity to explore culturally sensitive design riddled with tensions, such as the tension between correcting health ``misconceptions'' and respecting cultural or religious beliefs \cite{deva2025kya}. By partnering with a specific community and centering their perspectives, community-based research may drive designs that are more sensitive to sociohistorical context and intersectional identities \cite{harrington2019deconstructing,henriques2025feminist,kotturi2024sustaining,baseman2025we,alagh2025diaspora}. 

Others in HCI have taken community-engaged approaches to health technology design with Black or African American communities. O'Leary et al. partnered with church communities to explore the incorporation of spirituality within health technologies including a mobile health application \cite{o2022community} and a virtual agent \cite{o2020community}. Harrington et al. have engaged Black older adults in community centers and residential living facilities. Their findings reveal community desires to view ``health as a community practice'' with a focus on environment and fostering health activism \cite{harrington2019engaging}. When exploring health information seeking with a voice assistant, Harrington et al. found that their participants (Black older adults) largely believed voice assistants were not designed for them \cite{harrington2022s}. Some participants faced frequent communication breakdowns because they needed to ``code-switch'' to be understood by the agent. This suggests further marginalization of this community through the agent's design and a propagation of power dynamics associated with ``standard'' and ``non-standard'' English \cite{harrington2022s}. Baseman et al. leveraged a social-ecological approach to their partnership with community sites of Black older adults to uncover insights into mobile health for diabetes \cite{baseman2025we}. Finding experiences of medical mistrust within the community, they question how to acknowledge systemic inequity and infrastructural barriers within a health technology without disempowering community members. Empowerment is a key theme in community-based work with historically marginalized communities \cite{baseman2025we,harrington2019engaging,o2025empowering}, highlighting the value of culturally responsive health applications that make users feel validated \cite{o2025empowering}. 

This works offers, to the best of our knowledge, the first comm\-unity-based study in HCI to explicitly center medical mistrust. Further, our work deviates from an existing tradition within medical mistrust literature. Medical mistrust is often labeled as a ``cultural barrier,'' suggesting it is a flaw or inherent trait of marginalized communities \cite{jaiswal2019towards}. This perspective is harmful, utilizing a deficit-based lens \cite{tuck2009suspending}. It also shifts the burden of overcoming mistrust onto the very communities who are already navigating systemic exclusion and inequity, rather than holding institutions responsible for creating and sustaining these inequities \cite{brandon2005legacy}. Our study challenges this view by exploring how health technologies could be designed to acknowledge and validate mistrust where communities believe it is warranted based on lived experiences and sociohistorical context. This leads us to reflect on our positionality and the ethical tensions inherent in considering medical mistrust in health technology design (Section \ref{reflection}), and we conclude our work by detailing a set of seven design principles that emerge using the lens of Black Feminist Thought \cite{collins1990black,collins2000black} (Section \ref{framework_principles}). Driven by our reflexive themes, we propose this framework to guide the design of more ethical and equitable health technologies for communities with historically grounded medical mistrust.

\section{Methodology}

To explore community perspectives on doctors, the healthcare system, health AI, and medical mistrust, we conducted semi-structured interviews with 16 residents of two publicly subsidized apartment buildings. These community sites predominantly serve Black older adults with low income in a Southern U.S. state. Below we provide an overview of our partnerships with the apartment buildings and our study approach.

\subsection{Community Partnerships}
\label{method_partnerships}

Our team leveraged an ongoing partnership with two apartment buildings within a Southern U.S. city. At the time of the study, these partnerships were approximately 2 years and 3 years, respectively. As publicly subsidized housing under the city's housing authority, the apartments are designated for older adults (55 years or older) or people with disabilities, with an income below a percentage of the area's median income. The majority of residents are older adults with low income who identify as Black or African American. 

Our team takes a sustained engagement approach to these community partnerships. This is an approach that aims to foster more equitable engagements with under-resourced community sites via reframing, centering ethics and reflexivity, and engaging in sustained, community-driven outreach \cite{baseman2025we}. Our team works with site co-facilitators (including the elected presidents of both residents associations) to organize ongoing community outreach sessions. Weekly or monthly, the team visits the apartment buildings to assist residents with digital literacy and offer troubleshooting for their technological devices. Our motivation for leveraging a sustained engagement approach to community partnership is to ensure ethics and alignment to community values given the social distance between researchers and community members, to benefit the community through digital literacy support, and to build a relationship with the community  outside of the research study. This also aligns with a view of ``continuous and multiple engagements with communities and sites of research rather than a frame of giving back,'' as suggested by Bhan \cite{bhan2014moving}.

\subsection{Study Recruitment \& Participation}

Our interview study protocol was approved by our university's IRB. To ensure study suitability, the first interview at each apartment building was conducted with the president of the residents association. Data collected from the co-facilitators was included within the dataset, but the co-facilitators were encouraged to provide any feedback about questions they felt should not be asked to the residents at their building (e.g., if too sensitive, invasive, or confusing). We deferred to our co-facilitators on these judgments and in fact we did redesign the study by removing a survey (the Medical Mistrust Index \cite{laveist2009mistrust}) from one site's protocol after a co-facilitator felt the language was insensitive. Study recruitment was then conducted via printed flyers, which were created by the research team and distributed by our co-facilitators. 

Before taking part in the interviews, participants provided signed consent. The informed consent process included an explicit disclosure that the interviews would ``deal with sensitive topics, including your view of healthcare professionals and how they treat you and others in your community,'' and that participants could skip any question or end the interview at any time. Participants engaged in a 30- to 60- minute semi-structured interview, conducted by either the first or second author. The focus of the interview was to explore their perceptions of doctors and the healthcare system, and their views on medical mistrust. The last section of the interviews explored community perceptions and trust of an imagined computer vision-based system for self-monitoring diabetic foot ulcers. Throughout, the interviewers anchored the discussion by inviting participants to share stories and specific experiences to help us better understand the context and nuance of participant experiences. In addition, interviewers intentionally made a habit of repeating information back to participants in an attempt to ensure understanding to the best of our ability (e.g., ``I want to make sure I'm understanding. It sounds like you're saying that you... Do I have that right?'')  After the interview, we provided participants a \$20 gift card to a  grocery store chain. See Appendix \ref{appendix_protocol} for an overview and example questions from the semi-structured interview protocol. 

Interviews were conducted with 17 community members. The data from one participant, however, was unusable due to technical issues with a recording device. We were therefore left with 16 interviews, eight from each community site. Nearly all (n=15, 94\%) of our participants identified as Black, African American, and/or African. Twelve participants (75\%) identified as female, and the rest (n=4, 25\%) identified as male. Four participants were in their 40s or 50s (25\%), but the rest were in their 60s or 70s (n=12, 75\%). Fourteen (88\%) self-reported a diagnosis of type 2 diabetes and the other two participants identified as a caregiver of someone living with diabetes. Eleven participants (69\%) had completed at least some college, including five (31\%) who earned a bachelor's degree or higher.  Most (n=14, 88\%) self-reported having health insurance through Medicaid and/or Medicare. There was variety in the number of doctors participants routinely visited, ranging from just their primary care physician to a team of doctors. Most participants (n=12, 75\%), however, reported that they had visited a doctor within the last two months. Table \ref{tab:participants} provides an overview of participant demographics (see Appendix \ref{appendix_demographics}).

\subsection{Data Collection \& Analysis}

Interviews were audio recorded, automatically transcribed in Dovetail, and then manually reviewed to improve accuracy. 

The first and second author collaboratively conducted bottom-up, reflexive thematic analysis \cite{braun2019reflecting,braun2023toward}. While aiming to distill ``patterns of shared meaning'' \cite{braun2019reflecting}, we engaged in iterative reflection via discussions of positionality and our interpretations of the data. The transcripts were first open-coded in two batches during data collection. Each author created an initial proposed codebook based on their open-coding of the first batch of ten interviews. They then met to compare their codebooks and agree upon an initial set of themes. After open-coding the remaining interviews, the authors came to consensus on a final set of  reflexive themes related to participants'  mistrust or trust and how they navigate their health. An overview of themes from our work was disseminated back to community members via flyers (reviewed and approved by our co-facilitators). 

While our team has sustained partnerships with the community sites, we do not have the specific lived experiences of our participants. Our positionality as at least partially out-group researchers necessarily influences our interpretation of community members' experiences. We analyzed the interview transcripts with reflexive thematic analysis due to its emphasis on researcher subjectivity and interpretive influence \cite{braun2019reflecting,braun2021one}. Still, by distilling ``patterns of shared meaning'' \cite{braun2019reflecting} and choosing participant quotes, we are engaging in an extractive practice that portrays participant experience as deceptively shallow. We urge readers to consider that the participant quotes they will read in the next section are distillations of entire lived experiences. We have attempted to represent community members' sentiments as accurately as possible, and reframe our findings through Black feminist epistemology in Section \ref{framework_principles}.

\section{Findings}
\label{findings}

\begin{table*}[h!] 
\centering
\caption{Eight Themes Distilled from Reflexive Analysis}
\label{tab:themes} 
\renewcommand{\arraystretch}{1.2}
\setlength{\tabcolsep}{8pt}
\begin{tabular}{|p{16em}|p{25em}|}
\hline
\textbf{Reflexive Theme} & \textbf{Example Quote} \\ \hline

Doctors Vacationing to Hawaii: \newline Questioning Financial Motivations \newline & ``It shouldn't take that long for a doctor to figure something out [...] It sound like it's just all about the money.'' (P1) \\ \hline

Big Pharma and Agreeing to Disagree & ``The seniors in this building, I've known them to take anywhere from six to up to fifteen different pills. And I remember growing up, my grandmother never took all that stuff.'' (P5) \\ \hline

Teasing Apart the Accredited Truth \newline and the Embodied Truth & ``I'm telling them what's wrong with it. They give me [...] an x-ray. But then they still saying [...] nothing wrong. But I know something is wrong.'' (P1) \\ \hline

Always Look on the Bright Side & ``Even if I had a bad experience with a doctor, I'm still not going to focus on the bad experience. I'm still going to pull out the good I got.'' (P9) \\ \hline

``Rush You Out Like Cattle'': \newline Losing that Personal Touch \newline & ``Doctors used to actually care about the patient's well-being. Now they too busy. They got to play golf.'' (P12)  \\ \hline

Facing Bad Outcomes: \newline Malpractice or Victimism? & ``They hadn't went to the doctor. And everybody in the building, rather than saying that person didn't follow through on their health, they said the doctor killed [them].'' (P6) \\ \hline

Engagement and Agency as Drivers \newline of Appropriate Trust & ``Still highly mistrust the healthcare system. However, [I'm] building more trust because I'm taking more charge of what I need and want and what's right for me.'' (P7) \\ \hline

``Waiting on Aliens:'' Health AI is Only \newline As Good As the Intention Behind It \newline & ``It depends on what they're using [AI] for. You know, I really don't know if we either trust or mistrust, because it just \hspace{2em} depends'' (P13)  \\ \hline

\end{tabular}
\end{table*}

From semi-structured interviews with community members, we distilled eight reflexive themes (see Table \ref{tab:themes}) related to their experiences with healthcare professionals or institutions, and their perspectives on mistrust. The subsections below describe each theme in turn. 

\subsection{Doctors Vacationing to Hawaii: Questioning Financial Motivations}  
\label{findings_motivations}

A belief that some (or most) doctors are only interested in a paycheck contributed to medical mistrust in the community. P6 for example, recalled a doctor who ``didn't even touch my hand and he'll diagnose me with something,'' stating that ``I don't think he care. I think he's a money doctor. That's all he's about.'' P12 similarly suggested that some doctors are  ``more interested in paying off college loans. '' P7 further noted their belief that ``it just seems like everything is about money today. It's not about helping this one get better.'' Participants believed, for example, that this financial motivation led doctors to perform unnecessary tests: ``It shouldn't take that long for a doctor to figure something out [...] But then they started sending me the x-rays and giving me a cookie and some kind of liquid. It sound like it's just all about the money'' (P1).

Because of this perceived financial motivation, community members felt that the care you receive (and therefore the trust you should place in your doctor) is influenced by how much money you have. P13 expressed the belief that ``if I look like I come from the projects or I look like I'm raggedy, they not gonna probably give me the best treatment.'' P1 explicitly stated that ``I guess I trust them based on my ability to pay,'' humorously noting that ``because here in this country, it's about money,'' even ``Jack the Ripper'' could get adequate care if they had the funds. P5 also mentioned skepticism regarding cures for diseases such as cancer, noting that ``if they cured everybody, it would be a wrap. You got celebrities who've had AIDS for years, they're still living a happy life [...] because they have medicine for that. And those who can afford it, they get it.''

Financial motivations of doctors are therefore interwoven with another common source of medical mistrust: the pharmaceutical industry. P7 suggested that ``behind closed doors, the doctors are getting trips to Hawaii and stuff like that, if they pump these pills to the kids.'' We devote the next theme to exploring pharmaceuticals in more detail.

\subsection{Big Pharma and Agreeing to Disagree}
\label{findings_pharma}

Nearly all participants expressed mistrust of pharmaceutical companies, or at least skepticism regarding medications. Some participants, when asked about the source of medical mistrust, explicitly (and emphatically) stated that it stemmed from pharmaceutical companies. P5 described the pharmaceutical industry as a ``racket,'' and ``we be the guinea pigs.'' They noted their belief that ``the pharmaceutical companies have a [...] major part in overloading, especially seniors, with meds that they don't necessarily should or have to take. [...] The seniors in this building, I've known them to take anywhere from six to up to fifteen different pills. And I remember growing up, my grandmother never took all that stuff'' (P5). 

Even without these suspicions, other community members questioned the benefits and risks of medications. P10 lamented ``all these drugs out here and all these vitamins [...] and you have to wonder whether, is any of these things going to help you or you just taking them?'' Of course, being unhelpful is not the worst a medication can do when you consider potential side effects. P4 shared an anecdote when rather than ``diving more into trying to find out why things are not working for me,'' their doctors  took a trial and error approach, trying out a long line of medications. P4 said, however, that they ``always read the papers that they send with the medications,'' and shared their incredulousness and anger when they saw that the first listed side effect was that ``there could be \textit{death}.'' In different words, P13 noted that ``most of the time [the doctors] say they don't want to take [a medication]. Well, if it ain't good for you, why the hell is it good for me?'' Because of such skepticism, some community members do not follow recommended medication regimens. P5 shared that they ``stopped taking all that medicine, right. And I went to the doctors and [...] he was like, `Oh boy, you're doing great. I don't know what you're doing, but whatever you're doing, keep it up.' I stopped taking all the pills they were giving me, you see!''

Several community members spoke favorably of doctors who reduced their medications. For example, P6 contrasted their current doctor to a previous one, explaining, ``They're taking me off medicine. They're not working for the pharmacy, I noticed. [...]  I was taking like 10 pills a day. Now I'm down to 5. And I know I need them.'' In addition to this ``less is more'' principle, community members noted an appreciation of doctors who acknowledge a patient's preference for alternative or traditional medicine. P7 expressed that ``certain things you don't have to take pills for. And they don't teach us that,'' and P13 noted that ``everything we need to truly be healed to some extent, in my opinion, can be found in the garden.'' P13 further asserted that ``the bottom line is this. If I tell my doctor I don't really want to do pharmaceuticals, [...] I mean, that's a liability for him not to say, `Well, you know, we do suggest it.' [...] But he doesn't push it or force it or try to lie or try to manipulate me [...] because I'm like, I'll teach him something about Big Pharma.''

\subsection{Teasing Apart the Accredited Truth and the Embodied Truth}
\label{findings_truth_types}

Some community members did express strong trust in doctors, or at least a strong belief that they \textit{should} trust doctors. As P2 said, ``I mean, they the doctor. And you would think they would know what they talking about.'' P4 took pause in describing whether they trust doctors: ``Yes and no? Because we're supposed to, really, I think? [...] Because you look at them, they-- they're [with emphasis] \textit{doctors}, you know? And we just think [...] that they're supposed to know what they're doing.'' For many, this automatic trust was rooted in accreditation and education: ``I'm assuming they have their credentials. [...] I know initially whoever I'm going to, they went through some criteria to get to that point in their life'' (P3). P5 echoed that ``we're geared to trust them. I mean, you go to the hospital because of their knowledge, their medical knowledge, their schooling. [...] So you definitely go in trusting.'' 

Not everyone seemed to put this epistemological emphasis on accreditation and education, instead emphasizing their embodied experiences. P8 asserted that they trust doctors ``if they tell me the truth.'' When prompted further about how they determine if they're being told the truth, P8 responded: ``if they can help me. If [...] what they're telling me is true and if they [...] created something to give you a cure.'' It is not surprising that one would lose trust in doctors (or broader healthcare systems) if ``they ain't had nothing to help me yet [...] I mean, you can't help somebody in 76 years?'' (P8). This suggests that, for some, trust is less about the schooling behind medical knowledge, and more dependent on a doctor's ability to bring about bodily outcomes aligned with a patient's expectations. P1 similarly noted that they were put through a barrage of tests even though ``I'm telling them what's wrong with it. They give me [...] an x-ray. But then they still saying [...] nothing wrong. But I know something is wrong.'' P9 shared a story about their grandmother, who they learned ``was like a medicine woman. She'd go in the woods and pull some medicine, slap it on you, and you'd be well.'' P9 places less value on what is included in medical textbooks and instead ``trust[s] medicine as long as it's working for me.''

\subsection{Always Look on the Bright Side}  
\label{findings_bright_side}

Although most participants shared negative experiences or attitudes regarding health experiences, many also expressed a desire to keep optimistic. This was especially salient in regards to historical examples or keeping up with the news. P2 noted that, in their opinion, many older adults become less trusting as they get older, but they ``try to always think on the positive side.'' P15 expressed their dislike of the news, due to ``all this killing and all this, all that. I don't want to hear all that. I want to hear something good.'' When P3 could not think of any historical events related to medical mistrust, they explained that ``I don't know that because some things that are bad things, I try not to leave in the forefront of my memory.'' P9 ``look[s] for the good in everything'' and stated that ``even if I had a bad experience with a doctor, I'm still not going to focus on the bad experience. I'm still going to pull out the good I got.''

For some participants, this positivity was inherently tied to their religion or spirituality: They trust in God above any doctor, but believe that God put those doctors in their path. P12 did not stress about missing a doctor's appointment because ``if I choose to go to the doctor, I go. If I don't feel like going right now, I ain't going. [...] It's between me and God. I feel He'll take care of it.'' P13 identified as ``a positive, spiritual person'' who would not completely write off an ``evil'' doctor because ``God may be trying to get their heart to change. [...] Me personally, I look at everything, to some extent, from a spiritual standpoint. God could use that person. Even though they might be evil, they might help somebody.''

\subsection{``Rush You Out Like Cattle'': Losing that Personal Touch}
\label{findings_cattle_personal}

Accounts of previous experiences revealed how appointments felt brief and transactional, leaving little room for thoughtful care. P12 noted that things just aren't how they used to be, that ``doctors used to make house calls. Doctors used to actually care about the patient's well-being. Now they too busy. They got to play golf. `I got a six o'clock dinner reservation.'\thinspace'' Many participants shared this frustration. P4 remarked how healthcare organizations ``rush you in and rush you out like cattle.'' P16 noted that doctors ``spend, what, seven minutes with you and send you on your way? That's not cute. I need more time,'' stressing that doctors ``need to get more into their patients, get more understanding of what they're really, really going through.'' P3 agreed, noting that frequently, ``[I] take my 15 [minutes with the doctor] and I try to push it more than 15. [...] I try to stretch it.''

It follows that trust was built by doctors who intentionally take time with patients, who really listen with a personal touch. As P12 described: ``He sits there. He talks to me. [...] He listens. His nurse calls back to check on me. I trust him.'' Others also stressed the importance of a doctor circling back to check on something. This makes P4 ``feel really great that, you know, I'm being listened to. [...] And even when I bring up something, you know, she’ll say, ‘Okay, we're going to check on that. Let's just make sure that everything's okay with you.’\thinspace'' 

Community members especially value this extra time when it drives more patient-centered care.  P7 noted, for example, that some (but not all) patients without a high school diploma might have lower literacy, suggesting that doctors should take the time to ``know [their] audience, [...] to draw that out of them so you know what you're dealing with.'' P9 emphasized the value of forming a relationship and connection with their doctor: ``You don't seem to get enough attention [...]. I need somebody to support me and help me with this [diabetes] journey because I've been slipping, and I don't want to slip. You know, I want to master this. [....] When I slip, I need something to catch me.''

\subsection{Facing Bad Outcomes: Malpractice or Victimism?}
\label{findings_bad_outcomes}

As P10 noted, medical mistrust may exist in ``people who had an experience and something bad happened. They had a surgery, and it didn't come out the way it's supposed to come out.'' Community members differed, however, in where they placed accountability for negative clinical outcomes. For some, medical mistrust grew when they believed a doctor was negligent. Others believed that mistrust increases when a doctor gets blamed for something that was really the patient's fault (i.e., victimism). 

In this first perspective, community members believed doctors were negligent, incompetent, or at least should have done better. P1, for example, was dealing with an unexpected complication after an operation, and P14 lamented how clinic staff ``made a pincushion out of me'' when they ``couldn't find my veins.'' Some harbored negative sentiment towards doctors involved in a loved one's death. P6 noted that  “I think they could have did better with [a sibling's] health. [...] we still don't know what [they] died of. I didn't like the doctors, the medical community, or nothing. Nothing about it.'' P7 similarly described their late spouse's  diabetes-related amputations, saying ``we wasn't getting the right information until it was too late. You know, first the toe, then the foot, then the leg. [...] I just felt like we didn't get all the stuff we needed.''

Other community members held a contrary perspective: that medical mistrust arises unfairly due to placing blame on doctors. P9 asserted that ``the doctors are telling you what you need to do and you need to do it. And if you don't do it, you can't blame them for losing a damn foot.'' P15 echoed this sentiment, saying that others experience medical mistrust  ``cause they don't listen. [...] Somebody passes away, they be so sick, because they don't listen.'' P11 believed that mistrust could arise due to an assumption that a diagnosis was incorrect, expressing that ``I guess I didn't want to accept it when I was diagnosed with type 2 diabetes, but I thought maybe they had made a wrong diagnosis.'' P3 and P6 cautioned against placing blame on a doctor or assuming malpractice even when a loved one dies, as  ``[death] is a part of the word called L-I-F-E'' (P3). Especially living in a community of older adults, P6 highlighted that ``we're gonna die. [...] Like, two people died last week, but they were really ill. They hadn't went to the doctor. And everybody in the building, rather than saying that person didn't follow through on their health, they said the doctor killed [them].''

\subsection{Engagement and Agency as Drivers of Appropriate Trust}
\label{findings_agency_appropriate_trust}

Even beyond instances of negative clinical outcomes, accountability was a nuanced theme discussed by nearly all community members. Many felt that taking more personal accountability for your health builds medical trust and, more concisely, that a lack of \textit{appropriate trust} feeds mistrust. 

Taking a more active role in their care moved community members towards this desirable level of trust, e.g., by asking questions, ``go[ing] deep into what they saying that they're gonna do'' (P1), and doing your own research so you ``don't just believe everything they tell [you]'' (P12). P7 echoed that they ``still highly mistrust the healthcare system. However, [I'm] building more trust because I'm taking more charge of what I need and want and what's right for me.'' Others believed they had previously felt too much automatic trust of doctors, taking a hands-off approach to health until they learned that ``I had to be a part of it'' (P6), that they cannot simply ``put [...] [your health] in the hands of somebody else'' (P2). 

This agency was a crucial aspect in building medical trust. Participants who did not experience much mistrust stressed the ability to switch doctors or choose not to follow their recommendations. P13 and P16 expressed this walk-away power, that if ``they trying to do something weird like trying to get me to do something I don't want to do, [...] I'm not gonna deal with them.'' When faced with a doctor you do not trust, P16 stated that you should simply ``move around to the next one. If one ain't working for me, Imma move to the next one.'' Knowledge of this control decreased any ``concerns of [going to] a doctor, because I know that is an option. I can always go and change a doctor'' (P3).

Others expressed a need to also take an active role in their attitudes towards the healthcare system. P9 and P7 hinted at the idea of victimism, suggesting that those who experience medical mistrust may be to blame. P9 called mistrust ``a mindset,'' asserting that many with medical mistrust ``are stuck in their own beliefs'' that ``the system is against you.'' P7 had started to intentionally alter their viewpoint, saying that ``I need to grow and I need to learn. And if I'm being bitter and blaming others for everything, I can't fix the way I'm thinking.'' P9 and P7 therefore believed that it is partly the responsibility of those who experience medical mistrust to decrease this mistrust.

\subsection{``Waiting on Aliens:'' Health AI is Only As Good As the Intention Behind It}
\label{findings_intention}

When asked to imagine an AI-based system to help diabetes doctors, community members pointed out that these systems are not magic: ``Anything a computer does, a man told it to do. Simple. That's it.'' (P12). Because AI systems are programmed by humans and could be used in various ways, community trust in health AI was based on intention, motivation, and use: ``It depends on what they're using it for. You know, I really don't know if we either trust or mistrust, because it just depends'' (P13). 

P7 believed that health AI could decrease medical mistrust ``if it's done right. If it's done with good intentions.'' They further noted that ``if AI is used for good purposes, I would have a lot of trust in it [...] but I know somebody's going to start some evil mess with AI'' (P7). P13 similarly qualified their favorable view of AI, noting that ``the purpose of [AI] was to make life easier [...] so yeah, I love it, [...] as long as there's not any nefarious intent. Because [...] somebody might have an evil mind'' (P13). Others grounded their skepticism by questioning broad and uncritical applications of AI: ``I'm telling you, the AI can't do it all. They're trying to make it do it all, and there's going to be some problems down the road'' (P10). P8 also critiqued the larger commitments behind technological advancements, noting that although astronauts ``could go to the moon and back,'' doctors had not yet been able to assist P8 with their chronic health condition. They asserted that technology researchers should ``help people here! [...] Ain't no aliens out there told me nothing about no healing. Waiting on y'all aliens, then. Come on, bring us something to the Earth that can heal us'' (P8).

Community perceptions of health AI were also intertwined with perceptions of doctors. P15 believed that doctors are ``not there to hurt [people], they're here to help them'' and was therefore ``all for'' health AI (again, provided that ``nobody has ill intent to hurt someone else.'') Others suggested, however, that a doctor's use of AI would increase mistrust. They questioned why a doctor would need that assistance in the first place: ``What are they trusting a computer for, when you went to school to study all of this? Why you need a computer to tell you something?'' (P12). P10 echoed the sentiment, asserting that ``that's why you went to medical school. [...] You can't use AI for everything.'' Overall, community members perceived health AI based on the human motivations behind it, expressing optimism if used with good intent, but skepticism if applied nefariously or where it may not be appropriate.

\section{Discussion}
\label{discussion}

Recent work on trust in health technologies and institutional distrust demonstrates the influence sociohistorical context has on perceptions of technology. Leveraging community partnerships with older adults in publicly subsidized housing, our work explores historically grounded medical mistrust as an important consideration for the community-engaged design of health technologies. We conducted interviews with 16 Black older adults with two goals: We aimed to better understand community experiences of medical mistrust, as well as to further contextualize mistrust through a scenario of a health AI technology. 

We now unpack the reflexive themes detailed in Section \ref{findings}. We first discuss researcher positionality and ethical standpoint as critical factors that influence our engagements with under-resourced communities (Section \ref{reflection}). To reframe our findings, we then leverage Black Feminist Thought, an epistemological approach that centers the experiences and knowledge of Black women \cite{collins1990black,collins2000black}. Through the lens of structural, disciplinary, hegemonic, and interpersonal power, we suggest seven principles for advocacy and resistance through health self-management technology tailored to communities that experience medical mistrust (Section \ref{framework_principles}).

\subsection{Reflecting on Researcher Positionality and HCI's Role in Medical Mistrust}
\label{reflection}

We return to our team's positionality to better contextualize our findings and their implications. We leveraged multi-year partnerships, but the community sites are under-resourced and our participants are  members of intersectionally historically marginalized communities (e.g., Black women with low income). Our team is at least partially out-group researchers, with the privileges associated with working at an R1 institution\footnote{Notably, many of the universities designated as R1 practiced official or de facto segregation. In addition, of the currently 187 R1 universities in the United States, only one is a Historically Black College and University (HBCU).}. Further, medical mistrust has a long and complex history, with much of the blame falling on medical professionals, medical institutions, and the United States government. It may be reasonable to inquire, then, if HCI researchers need  concern themselves with medical mistrust. After all, we simply research and design technology.

Our team is fueled by visions of equity and ultimately aim to support these communities; however, we entered this work with our own values and lived experiences. We grappled with questions of appropriate responsibility to the community, of what role we should be play given what we had learned through our work. We believe we may spend our entire careers (or lives) exploring these questions, but our team discussed at least three potential ethical positionalities.

First, HCI researchers could \textbf{stand for} their own commitments and values. This may be at odds with a human-centered design approach, which places an emphasis on centering  users and sensitivity to context \cite{sambasivan2009human}, in part to avoid creating technologies that are unusable, useless, or unethical \cite{veinot2018good,sharp2019interaction}. A ``stand for'' approach would diminish the voices of community members, instead holding researcher beliefs paramount. This may reduce the agency granted to community members and risk a mismatch between researcher assumptions and the community. For example, this approach is likely to misunderstand or ignore  medical mistrust, given the literature's over-emphasis on the Tuskegee Syphilis Study  \cite{jaiswal2019towards}. Consider, however, a researcher with a value of mitigating health disparities and commitment to conventional medical knowledge. ``Standing for'' these values would allow them to design technologies that minimize health misinformation and decrease  mistrust, perhaps improving health outcomes (as is suggested by a wide body of literature \cite{souvatzi2024trust,shukla2025medical,williamson2018systematic}). This provides a path forward for technological opportunities to mitigate disparities that community members may not understand due to lower technological (e.g., predictive machine learning algorithms) or health literacy (e.g., vaccines) \cite{barros2021circumspect,bondu2024accessible}; however, it may come at the cost of invalidating voices of skepticism and historically situated mistrust.

Second, HCI researchers could \textbf{stand between} community members and institutions that propagate inequities. HCI researchers might have a responsibility to spread awareness about historical harms (i.e., educating researchers, policy makers, etc.) and to increase health and technological literacy. Much prior literature advocates for bolstering the literacy of under-resourced community members or older adults through tailored educational initiatives \cite{bondu2024accessible,tang2025ai,evans2021designing,zhai2025hear}. Literacy, however, should not be viewed as a silver bullet: It implicitly places the blame for inequities on those facing health disparities and increases the burden placed upon them. This drives an alternative version of ``standing between.'' While some studies document community skepticism of novel health technologies \cite{barros2021circumspect,harrington2023trust}, other work on the machine heuristic suggests an overall tendency to view machines as more objective and reliable than humans \cite{sundar2019machine,gambino2019digital}. Health technology designers could attempt to better understand medical mistrust in order to mimic this protective response with our designs. This attitude of ``protection'' may be well-intentioned, and does not evidently place additional burden on community members, but it is paternalistic and, again, reduces the agency of community members \cite{kender2025social}. 

And third, HCI researchers could \textbf{``stand with''} community members. Kim TallBear's formulation of ``standing with'' a community breaks down the false boundary between researcher and researched common in some forms of knowledge production \cite{tallbear2014standing}. To ``stand with'' community members, you must approach them ready to be changed by the partnership, to realign your commitments, values, beliefs, and epistemological alignments. As community engagements are bi-directional, this requires not only deep reflexive engagement, relational accountability, and commitment of the researchers but also of the community itself. Gaining deeper knowledge of medical mistrust and health experiences should change the ways we design and approach research moving forward. It is complicated, however, by the diversity of community experiences. Not only did the community defy a monolith narrative, but members occasionally held contradictory opinions, e.g., valuing the accredited versus the embodied truth (Section \ref{findings_truth_types}), viewing negative health outcomes as malpractice or as victimism (Section \ref{findings_bad_outcomes}). ``Standing with'' the community, then, is a practice not only of care and reflection, but also of nuance and pluralism. 

Our team finds that these three standpoints are dynamic. They  have strengths and weaknesses, and we move across them fluidly depending on contextual nuances. Notably, a trade-off appears between community agency and the amount of burden placed on community members. We articulate these standpoints to: 1) urge readers to reflect on their own values within this negotiation, and 2) to better ground Section \ref{framework_principles}, as positionality and epistemology are deeply linked.

\subsection{Designing Health Self-Management Technologies with Medical Mistrust: Opportunities for Resistance Through Black Feminist Thought}
\label{framework_principles}

\begin{figure*}
    \centering
    \includegraphics[width=.6\linewidth] {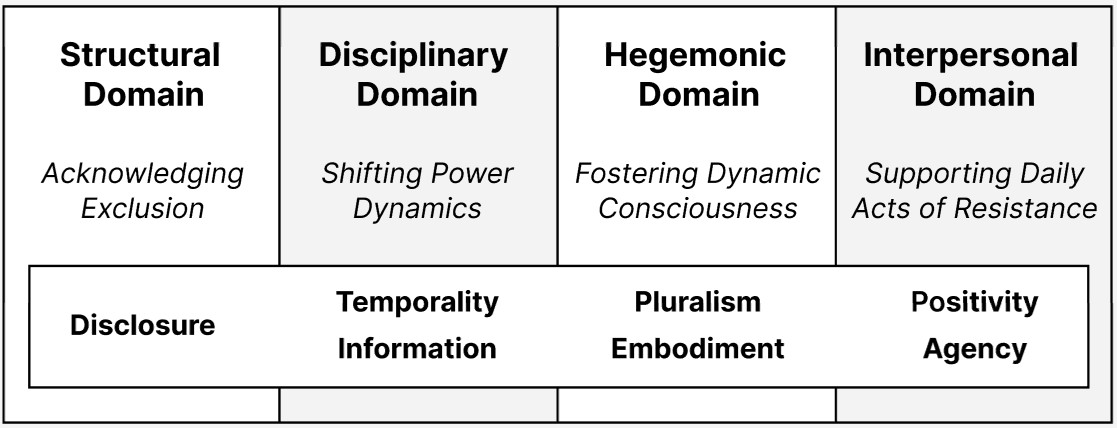}
    \caption{Seven principles for designing health self-management technologies tailored to communities that may experience medical mistrust: disclosure, temporality, information, pluralism, embodiment, positivity, and agency}
    \label{fig:principles}
    \Description{The figure shows four columns, associated with sections 6.2.1 through 6.2.4. At the bottom of each column, the associated design principle is listed. The first column reads ``Structural Domain'' and ``Acknowledging Exclusion,'' and the associated design principle is ``Disclosure.'' The second column reads ``Disciplinary Domain'' and ``Shifting Power Dynamics,'' and the associated design principles are ``Temporality'' and ``Information.'' The third column reads ``Hegemonic Domain'' and ``Fostering Dynamic Consciousness,'' and the associated design principles are ``Pluralism'' and ``Embodiment.'' Finally, the fourth column reads ``Interpersonal Domain'' and ``Supporting Daily Acts of Resistance,'' and the associated design principles are ``Positivity'' and ``Agency.'' }
\end{figure*}

Despite broad work in community-based and culturally sensitive health technologies, medical mistrust has not yet received much attention in HCI. Previous findings suggest that medical mistrust may translate into skepticism of health technologies \cite{harrington2023trust,lee2021included}, but better understandings of health experiences currently underrepresented in the health technology literature reveal potential paths forward for empowerment, resistance, and advocacy.

Given the importance of intention in health technology design (Section \ref{findings_intention}), we argue that continuing a tradition of HCI health literature that does not intentionally name and center historically grounded medical mistrust risks propagating ignorance. Epistemologies of ignorance assert that ignorance (a lack of knowledge or “unknowledge”) is embedded in social and power structures, and plays a functional role in knowledge production \cite{sullivan2007introduction}. Ignorance can be actively produced and its maintenance underpins racism and systemic inequities \cite{mills2007white}. By selectively not-knowing community perspectives and lived experiences, we produce knowledge that views the health-related experiences of white, middle- to upper- class individuals as the norm \cite{bowleg2017towards}. This risks maintaining the systems of health inequity that we aim to dismantle. 

Epistemology can also be used as a tool of resistance and empowerment \cite{sullivan2007introduction}. Black Feminist Thought (BFT) as described by Patricia Hill Collins \cite{collins1990black,collins2000black}, for example, critiques dominant epistemologies that exclude or undervalue the unique standpoint of Black women. Instead, BFT views Black women as agents of knowledge and centers their experiences. BFT has been previously utilized in HCI to bolster voices typically pushed to the margins in technology design, including those of Black women (e.g., in game design \cite{rankin2020seat} and their experiences on social media \cite{musgrave2022experiences}) and Black youth (e.g., for the design of conversational agents \cite{rankin2021resisting} and speculative responsible AI \cite{tanksley2025ethics}). A central concept of BFT is the Matrix of Domination, a framework that describes ``interlocking systems of oppression'' along axes of identity including (but not limited to) race, gender, socioeconomic status, and age \cite{collins1990black}. These systems of power function in four domains: structural, disciplinary, hegemonic, and interpersonal. In each, there also exist opportunities for resistance and activism. The Matrix of Domination, too, has been applied in HCI: Erete et al. have analyzed the role of power and opportunities for resistance in the technology design process \cite{erete2021can,erete2023method} and design collaborations \cite{erete2025towards}. 

In the rest of this section, we reframe our findings through each domain (structural, disciplinary, hegemonic, and interpersonal) to 1) ground suggested avenues for health technology designers to support community empowerment, and 2) to drive the emergence of a framework of seven design principles (see Figure \ref{fig:principles}).  In building this framework, we were also inspired in part by Bardzell’s highly influential conception of Feminist HCI \cite{bardzell2010feminist}. As Bardzell suggested for her six "qualities of feminist interaction” \cite{bardzell2010feminist},  we suggest that the following principles for health technology design occur in critical mass.

\subsubsection{Structural Domain: Acknowledging Exclusion Through Disclosure}

The structural domain officially organizes power through laws, policies, and other social organization that formalizes exclusion and inequity. Examples include institutions that have denied job opportunities, voting rights, and adequate health care \cite{collins1990black,collins2000black}. Our findings scratch the surface of the interlocking systems that exclude and underrepresent Black older adults with low income. In the health domain, community members noted their belief that the care you receive is based on how much money you have (Section \ref{findings_motivations}), pointing to the intersection of race and income. Further, community members greatly cared about the intention of programmers and doctors using health AI systems (Section \ref{findings_intention}). In the domain of technology, too, communities of color, women, and those with low income have historically been excluded and continue to be underrepresented in science and technology \cite{ovink2024figuring,rankin2020intersectional}. We suggest that researchers advocate against an epistemology of ignorance that denies structural racism and systemic inequities. A first step is to center the lived experiences and knowledge of those historically excluded and/or intersectionally marginalized. 

\textbf{Disclosure.} Our findings suggest the need to practice transparency through \textcolor{black} honest disclosure of the motivations behind health technologies. These designs should make immediately evident who is behind the technology/research and what their intentions are. Inspired by ethical machine learning model cards \cite{mitchell2019model}, future researchers could take a participatory community-based approach to create a standardized form through which health technology designers could disclose these factors. What are the history and commitments of the institutions funding the technology? Who benefits from its use and how does that impact the recommendations provided? One health self-management technology may seek to dispel ``misconceptions’’ about health and proactively urge users towards medications. Another, funded by those practicing historical revisionism, may spread disinformation regarding mistrust and health injustices. A corporation may intend to increase use of their product and perhaps sell health data to third parties. Other groups could design a self-management technology to bolster community empowerment via information about health inequities and alternative medicine. By transparently disclosing the motivations behind health technologies we have the opportunity to acknowledge historical exclusion and clarify our own intentions.

\subsubsection{Disciplinary Domain: Shifting Power Dynamics Through Temporality and Information}

Community member stories described ``money doctors'' (Section \ref{findings_motivations}) who rush their patients in and out (Section \ref{findings_cattle_personal}). Further, many participants expressed that they were not given the health information they needed, and questioned whether bad outcomes were evidence of doctor malpractice (Section \ref{findings_bad_outcomes}). This echoes previous findings emphasizing the importance of patient-clinician relationships for building trust \cite{benkert2019ubiquitous,souvatzi2024trust}, as a personal bond and open communication can make patients feel like ``more than just a number'' \cite{gagnon2025characteristics}. Although community members officially have access to health care, the incentive structures and organization of these institutions still find ways to restrict access to care and necessary health information. In the disciplinary domain, power is maintained not by strict exclusionary laws or policies, but rather through organizational control (e.g., bureaucracy, surveillance) \cite{collins1990black,collins2000black}. There exist opportunities to design health technologies to disrupt these power relationships, both in patient-facing and clinician-facing technologies. 

\textbf{Temporality.} In Section \ref{findings_cattle_personal}, we describe community members' desire for doctors who intentionally practice patient-centered care, rather than rushing through appointments. This suggests a quality of temporality in feeling valued, cared for, and validated. Health technologies could aim to aid users in self-advocacy to ``try to stretch'' (P3) their time during clinical interactions, or at least to make the short time they do have with doctors feel more meaningful and personal. 

\textbf{Information.} Timely health information (Section \ref{findings_bad_outcomes}) is especially important for individuals with medical mistrust who are concerned that health care workers are hiding information from them \cite{cueva2024medical}. Informational support, paired with clear and proactive expectations management, may help community members distinguish what is within their power, what is  within their doctor’s power, what is a collaborative effort, and what is uncontrollable. Health technologies might also be designed to support disciplinary resistance of those who can disrupt the power relations maintained by healthcare organizations. Consider clinician-facing technologies that provide more avenues for health professionals to act in solidarity with their patients, for example for nurses to directly send patients additional information regarding their procedures or potential malpractice concerns.

\subsubsection{Hegemonic Domain: Fostering Dynamic Consciousness Through Pluralism and Embodiment}

The hegemonic domain attempts to validate oppression through ideology and what is seen as ``common sense'' \cite{collins1990black,collins2000black}. As Patricia Hill Collins stated, ``the hegemonic domain becomes a critical site for not just fending off hegemonic ideas from dominant culture, but in crafting counter-hegemonic knowledge that fosters changed consciousness'' \cite{collins2000black}. In our findings, community members expressed (at least) two key health perspectives that do not align with the current hegemonic view of health in the United States. First, many community members expressed mistrust of doctors and pharmaceutical companies, instead showing interest in alternative medicine and natural remedies (Section \ref{findings_pharma}). This follows prior findings that those with medical mistrust may be suspicious about medical treatments and side effects, and therefore may prefer natural or traditional medicine \cite{shukla2025medical,gagnon2025characteristics,cueva2024medical}. Second, some viewed the ``truth’’ of their health as inherently linked to their experiences and physical outcomes, rather than basing their trust in a doctor's professional education (Section \ref{findings_truth_types}).

\textbf{Pluralism.} Defined by Bardzell as designs ``that resist any single, totalizing, or universal point of view’’ \cite{bardzell2010feminist}, we must find ways to support pluralism in health technologies. Rather than only providing conventional medical knowledge, we must encode different perspectives including respect for alternative or traditional medicine, transparency and validation regarding side effects, and an understanding of the historical context of medical experimentation. This pluralism would better account for the diversity of experience not only in health, but more generally in the context of protective responses to oppression \cite{collins2000black}. We pull from Haraway’s view of situated knowledge \cite{haraway1988situated} when we assert that we must trust individuals about their own lived experiences and grant them agency to make their own choices, especially when institutions of power have continually attempted to dampen that choice. A view from a body as opposed to a view from above, as argued by Haraway \cite{haraway1988situated}, positions knowledge not only as situated but also as embodied. 

\textbf{Embodiment.} Especially for those who have been systemically abused or ignored by medical systems, health technology designers should consider that what is right cannot be disentangled from what one's body tells them (i.e., experiential and physical outcomes). This motivates an emphasis on embodiment: symptom journaling, validation of users who feel that something is wrong even when tests have not yet concurred, and sensitivity to the frustration of long-standing symptoms that have not been effectively treated. An embodied understanding of health could support counter-hegemonic knowledge of health, rather than frustrating and further marginalizing some community members. Further, this could support the growth of ``embodied health movements,'' in which those living with an illness or disability drive social change by challenging ignorance within conventional medical science  \cite{whooley2021uncertain,brown2004embodied}.

\subsubsection{Interpersonal Domain: Supporting Daily Acts of Resistance Through Agency and Positivity}

Focusing on only the structural, disciplinary, and hegemonic domains can overlook the ways in which community members actively resist and practice activism in their daily lives. This interpersonal domain influences individuals' experiences and social norms \cite{collins1990black,collins2000black}. There are a number of ways community members communicated their daily acts of resistance, and therefore opportunities for health self-management technology to further support them.

\textbf{Agency.} Community members viewed agency and control over their health as a vital tool in building a desirable level of medical trust (Section \ref{findings_agency_appropriate_trust}). Regardless of whether or not they experience mistrust, community members felt it was their responsibility to take a more active role in their health to ultimately receive better care. Doctor shopping and researching questions online, for example, are known strategies individuals with medical mistrust may leverage to empower themselves \cite{gagnon2025characteristics}. Previous work has also suggested that mobile health (mHealth) and similar systems may facilitate empowerment in care \cite{hamberger2022interaction,baseman2025we,ng2019provider}. We note, however, that these systems need not do this passively, i.e., simply by existing in the hands of users. Technologies may be developed to actively bolster agency \cite{bircanin2025beyond}. For example, systems might intentionally allow for users to better interrogate doctors and make their own decisions. 

\textbf{Positivity.} Further, our work was partially motivated by the knowledge that health AI systems may perform worse for people of color (Section \ref{[project_history]}). We assumed that increased historical awareness (e.g., of experimentation on Black or African Americans) would empower critical health decision-making. Community members did not identify many historical injustices related to medical mistrust, however, and they highly valued positivity and optimism (Section \ref{findings_bright_side}). This led our team to reframe our view of historical awareness: An avoidance of additional emotional burden is entirely reasonable for communities already disparately impacted. When engaging users in mundane health tasks or when reporting bad news, health technologies should aim to empower users with the necessary information while keeping it light and positive. Dispelling a damage-centered view \cite{harrington2020forgotten}, self-management technologies should allow community members to paint their own narratives, including those of resistance, survival, and empowerment. 

\subsection{Limitations and Future Work}

Our community-engaged exploration of medical mistrust involved partnerships with only two community sites, which predominantly serve Black older adults with low income living within one city in the United States. We therefore do not aim to provide findings that are generalizable \cite{gaver2014science}, instead providing insights that \textit{may} be transferable \cite{soden2024evaluating} to other communities that experience historically grounded medical mistrust (e.g., other communities of color, queer communities). This transferability would have to be confirmed by  future research, given the historical specificity of the community in this study. Still, we believe the implications of our work and our proposed design principles offer a valuable starting point for reflexive discussions about ethics within HCI health research. We look forward to future iterations and insights on our proposed design principles.

Future steps also include the direct involvement of community members in efforts to iterate on proposed design considerations and collaboratively design associated health self-management technologies. These efforts would benefit from including local health equity experts and heavily emphasizing  community empowerment, shared governance, community priorities and values, and sustainability of any intervention.

An important limitation of our work is that all of our participants are residents of publicly subsidized housing. They likely have the resources and skills necessary to navigate complex institutional systems and processes, e.g., securing enrollment in these apartments. Our participant pool may therefore have excluded those who are more isolated, marginalized, and/or have higher distrust (or mistrust) of institutions funded by the state.

In addition, we note that the semi-structured interviews included sensitive topics, including views of the healthcare system and previous experiences with healthcare professionals. While participants were told that they could skip any question and end the interview at any time, this may have deterred community members who did not wish to disclose sensitive information to researchers. Even for participants who chose to participate, it is likely that some refrained from sharing certain personal experiences, views, or insights. In fact, multiple participants did choose to ``pass'' on a question, although no participants requested to entirely end an interview early.

\section{Conclusion}

Medical mistrust is a rational, protective response driven by a long history and current state of inequity, injustice, and harm \cite{benkert2019ubiquitous}.  Despite a growing emphasis on the community-based and culturally-sensitive design of health technologies, the HCI community has largely overlooked medical mistrust as a key aspect of health practices and perspectives of health technologies within Black or African American communities. We offer, to our knowledge, the first community-engaged study to specifically concentrate on medical mistrust within human-centered computing. We leveraged our multi-year partnerships with two publicly subsidized apartment buildings to center the lived experiences of Black older adults with low income in the Southern United States. Our findings describe community perspectives on the healthcare system and medical mistrust, including skepticism regarding doctors' financial motivations, a negotiation between trusting accreditation versus embodied experiences, the value of taking more agency over one's health, and the importance of the intention behind health AI. We provide a reflective exercise for researchers to consider their positionality as they dynamically ``stand for'' their values, ``stand between'' historically marginalized communities and institutions propagating inequities, and ``stand with'' \cite{tallbear2014standing} community members. By reframing our findings through Black Feminist Thought \cite{collins1990black,collins2000black}, we propose a framework of seven principles to guide the design of more intentional, ethical, and equitable health self-management technologies. It is our responsibility as health technology designers to be aware of medical mistrust and its historical context, to consider the lived experiences of those in historically marginalized communities, and to support them via acts of advocacy and resistance. To this end, we suggest centering disclosure, temporality, information, pluralism, embodiment, positivity, and agency.

\begin{acks}
We would like to thank Nathaniel Swinger and Kefan Xu for assistance with data collection, and Krishna Ravishankar for research insights. This work was funded by the American Diabetes
Association Grant 11-22-ICTSHD-09.
\end{acks}

\bibliographystyle{ACM-Reference-Format}
\bibliography{10_references}

\appendix

\newpage

\section{Overview of Interview Protocol}
\label{appendix_protocol}

Below we provide a high-level overview of our interview protocol including a \textbf{sample sub-set} of the questions asked to participants. Each participant interview covered five subsections (A.1 through A.5). Almost all of the participants were asked all of the bulleted questions listed below (i.e., first level of indentation) except due to constraints such as time or participant comfort. Questions following a dash (i.e., second level of indentation) are examples of suggested probing questions included in the interview guide. 

During the informed consent process, the study motivation was described as follows: ``The purpose of this study is to learn more about your experiences with health care, to help us design a tool to help support diabetes self-management.'' Participants were told that they would be asked about their ``thoughts on health care and doctors,'' their ``experience with health care technologies and artificial intelligence,'' ``how [they] would or wouldn’t use them to support [their] journey with diabetes,'' and their ``thoughts about the role [health AI] could play in diabetes management.'' 

\subsection{Background}

\begin{itemize}
    \item Do you have a diagnosis of diabetes? 
    \item Do you have any family members or friends that have diabetes? 
    \item When were you diagnosed with diabetes? 
    \item Which doctors do you see regularly (e.g., primary care physician, podiatrist, cardiologist)?
    \begin{itemize}
        \item How often do you go to see these doctors? 
        \item When was the last time you went to a doctor?
    \end{itemize}
\end{itemize}

\subsection{Perception of Doctors and Healthcare}

\begin{itemize}
    \item How would you describe your experiences with doctors?
    \begin{itemize}
        \item Do you trust your doctors? 
        \item What does trusting a doctor look like to you? Give me an example.
        \item When you meet a new doctor for the first time, do you trust them? Why?
        \item Can you tell me about a time when a previous experience with doctors or the healthcare system influenced your trust? 
    \end{itemize}
    \item How would you describe your experiences with the healthcare system more generally?
    \begin{itemize}
        \item Do you feel like the healthcare system has your best interests at heart? Why or why not?
    \end{itemize}
    \item We’re interested in something called ``medical mistrust.'' This is a belief that healthcare providers, institutions, or systems are acting against your best interest or well-being. Would you say you experience medical mistrust? 
    \begin{itemize}
        \item Have you always felt this way or have your feelings changed over time?  
        \item Why do you think other people might (mis)trust doctors or the healthcare system more/less than you do?  \\ \\ \\ \\
    \end{itemize}
\end{itemize}

\subsection{Causes of Medical Mistrust}

\begin{itemize}
    \item Where do you think medical mistrust comes from?
    \begin{itemize}
        \item Do people experience medical mistrust because of individual doctors? 
        \item Is there a pattern in hospitals/clinics in different areas? 
    \end{itemize}
    \item Do any historical events come to mind when you think about medical mistrust? 
    \begin{itemize}
        \item Do you think these events affect people's views today? 
    \end{itemize}
\end{itemize}

\subsection{Navigating Medical Mistrust \& Ecological Impacts}

\begin{itemize}
    \item How does your trust in doctors and the healthcare system influence how you choose to get care? 
    \begin{itemize}
        \item (If trust was low/medium) What do you do to make sure you're still getting the care you need, given your mistrust?
        \item (If trust was high) If you didn’t trust doctors or the healthcare system, what would do to make sure you're still getting the care you need? 
    \end{itemize}
    \item Do you think doctors consider a person's background, identity, or financial situation while treating them? 
\end{itemize}

\subsection{Medical Mistrust \& AI for Health}

\begin{itemize}
    \item Have you had a diabetic foot ulcer or any other foot complications due to diabetes?
    \begin{itemize}
        \item If yes, how did the doctor examine it and what was your experience like? 
    \end{itemize}
    \item Imagine a computer program that looks at pictures of feet and uses AI to help doctors find foot ulcers. What do you think about a system like that? 
    \item Where do you fall on a line between full mistrust (0) and full trust (10) of a computer program that uses AI to find foot ulcers? 
    \begin{itemize}
        \item Could this AI system impact your doctors’ ability to work towards your best interest? How?
    \end{itemize}
    \item How do you think a machine would go about identifying ulcers? 
    \begin{itemize}
        \item How would the machine decide if there is an ulcer or not? What factors would affect this decision? 
    \end{itemize}
    \item Overall, how do you think AI systems like this could affect people’s medical trust or mistrust?
\end{itemize}

\newpage

\section{Participant Demographics}
\label{appendix_demographics}

\begin{table} [h!]
\caption{Participant Overview (N=16)}
\label{tab:participants}
\renewcommand{\arraystretch}{1.2}
\begin{tabular}{ | p{0.07cm} l l | } 

\hline  \textbf{Demographic} &  &  \textbf{N (\%)} \\ \hline

\multicolumn{3}{|l|}{Gender} \\ \hline
 & Female & 12 (75\%) \\ \hline
 & Male & 4 (25\%)  \\ \hline
 
\multicolumn{3}{|l|}{Race/Ethnicity} \\ \hline
 & Black, African American, and/or African & 15 (94\%) \\ \hline
 & White & 1 (6\%)  \\ \hline
 
\multicolumn{3}{|l|}{Age}\\ \hline
 & 40-49 & 1 (6\%) \\ \hline
 & 50-59 & 3 (19\%)  \\ \hline
 & 60-69 & 10 (63\%) \\ \hline
 & 70-79 & 2 (13\%)  \\ \hline

\multicolumn{3}{|l|}{Highest Level of Education} \\ \hline
 & Graduated high school, 12th grade, or less & 5 (31\%) \\ \hline
 & Some college, no degree & 6 (38\%)  \\ \hline
 & Bachelor's degree & 4 (25\%) \\ \hline
 & Post-graduate degree & 1 (6\%) \\ \hline
 
\multicolumn{3}{|l|}{Health Insurance Coverage } \\ \hline
 & Medicare \& Medicaid & 6 (38\%) \\ \hline
 & Medicare only & 5 (31\%) \\ \hline
 & Medicaid only & 3 (19\%)  \\ \hline
 & No Response or ``Other'' & 2 (13\%) \\ \hline

\multicolumn{3}{|l|}{Last Visit to a Doctor} \\ \hline
 & Within the Last Week & 6 (38\%) \\ \hline
 & Within the Last Two Months & 6 (38\%) \\ \hline
 & Within the Last Four Months & 1 (6\%)  \\ \hline
 & Not Reported & 3 (19\%) \\ \hline

\end{tabular}
\end{table}


\end{document}